\documentclass[twocolumn,showpacs,preprintnumbers,amsmath,amssymb]{revtex4}

\usepackage{graphicx}
\usepackage{dcolumn}
\usepackage{bm}

\begin{document}


\title{Monovalent counterion distributions at highly charged water interfaces: Proton-transfer and
Poisson-Boltzmann theory }

\author{Wei Bu, David Vaknin, and Alex Travesset}
\affiliation{Ames Laboratory, and Department of Physics and Astronomy, Iowa State University, Ames, Iowa 50011}
\date{\today}

\begin{abstract}
Surface sensitive synchrotron-X-ray scattering studies reveal the distributions of monovalent ions next to highly charged interfaces. A lipid phosphate (dihexadecyl hydrogen-phosphate) was spread as a monolayer at the air-water interface, containing CsI at various concentrations. Using anomalous reflectivity off and at the $L_3$ Cs$^+$ resonance, we provide, for the first time, spatial counterion distributions (Cs$^+$) next to the negatively charged interface over a wide range of ionic concentrations.  We argue that at low salt concentrations and for pure water the enhanced concentration of hydroniums H$_3$O$^+$ at the interface leads to proton-transfer back to the phosphate group by a high contact-potential, whereas high salt concentrations lower the contact-potential resulting in proton-release and increased surface charge-density.  The experimental ionic distributions are in excellent agreement with a renormalized-surface-charge Poisson-Boltzmann theory without fitting parameters or additional assumptions.
\end{abstract}

\maketitle

The electrostatics of aqueous solutions is a rich and fascinating topic full of unexpected and counterintuitive phenomena that still presents noteworthy challenges, both theoretically and experimentally, vital for a complete understanding of the physics of biological systems\cite{Grosberg2002}. In recent years, there has been ample theoretical activity aimed at determining ion distributions next to highly charged interfaces. It has been predicted that multivalent ions become strongly correlated next to the interface \cite{Shklovskii1999,Moreira2001,Burak2004}, thus invalidating the traditional Poisson-Boltzmann (PB) theory\cite{Safran1994}. Direct experimental verification for these correlations has been almost non-existent, and only recently is gradually emerging\cite{Bedzyk90,Angelini2003,Vaknin2003,Bestman2004}. Although correlations among monovalent ions are unimportant, the distribution of monovalent ions at highly charged surfaces has also been controversial.  In the context of hydration forces\cite{Leikin1993} recent theories seem to favor interfacial restructuring of water\cite{Manciu2004,Faraudo2005} leading to ion distributions that may significantly differ from simple PB theory. First-principle predictions of surface-tension isotherms of surfactants assume the existence of a relatively large Stern layer with a dielectric constant lower than that of pure water\cite{Proesser2001}. Furthermore, detailed theoretical and numerical studies have shown the importance of including other effects\cite{Torrie1982}.  To settle these issues, it is essential to obtain precise experimental ion distributions, including distant points from the interface. This is also a necessary step for an unambiguous understanding of those electrostatic effects that differentiate monovalent from multivalent ions.

In this Letter, we report on experimentally determined monovalent ion-distributions at highly charged interfaces, with an effective surface charge density in the $\sigma_0\approx 0.08-0.4 \frac{C}{m^2}$ range (molecular area of $40-180 \AA^2$). Although our main interest in this study is in the context of biological physics, its relevance extends far beyond that to basic aspects of intermolecular forces, electrochemistry and possibly to plasma physics.

Poisson-Boltzmann (PB) yields ion distributions via two characteristic lengths, the Gouy-Chapmann length $\lambda_{GC}=k_BT\varepsilon/2\pi\sigma_0e$, and the Debye screening length $\lambda_D=\sqrt{\frac{\varepsilon k_BT}{8\pi e^2 n_b}}$, where $T$ is temperature, $k_B$ is the Boltzmann's constant, $\varepsilon$ is the static dielectric constant, and $n_b$ is the bulk salt concentration. In the high charge limit, i.e., $\lambda_{GC}/\lambda_D<<1$, PB theory predicts the distributions next to the charged interface are practically independent of bulk concentration for at least the first $10$\AA (for $10^{-5} \leqslant n_b \leqslant 0.1$M).

Herein, we point to a well understood, but frequently overlooked issue concerning charged interfaces at aqueous solutions. Nearly all biologically relevant molecules, including proteins, DNA, and many phospholipids become negatively charged by proton-release, the efficiency of which is given by $\alpha = 1/[1+10^{(pKa-pH)}]$.   Usually, the pKa $<$ pH and almost all protons are dissociated for neutral pH $ \sim$ 7 ($\alpha \approx 1$). However, when such molecules form an interface, in particular a planar one, hydronium concentration becomes significantly higher than bulk at that interface, leading to a lower interfacial pH and to proton-transfer back to the interfacial molecules. Thus, the net surface-charge is reduced.  Within PB the enhancement  is expressed quantitatively by the Boltzmann factor $\text{exp}(-\frac{e \psi(0)}{k_BT})$, where $\psi(0)$ is the contact value potential.  The effective charge at the interface is renormalized as follows\cite{Israelachvili2000},
\begin{equation}
\label{sigma} \sigma_r= \frac{\sigma_0}{1+10^{(pK_a-pH)}\text{e}^{\frac{-{e\psi(0)}}{{k_BT}}}}.
\end{equation}
The potential at the interface, $\psi(0)$, which can be influenced by ions in solution, is determined self consistently from the boundary-condition-equation $\sinh(\phi_{0}/2)=-\lambda_{D}/\lambda^{\prime}_{GC}(\phi_0)$, where $\phi_0\equiv e\psi(0)/k_BT$, and a renormalized Gouy-Chapman length $\lambda_{GC}^{\prime}= k_{B}T\epsilon/2\pi\sigma_{r}e$, equivalent to the Grahame equation. The counterion distribution is given by the Poisson-Boltzmann with a renormalized Gouy-Chapman length $\lambda_{GC}^{\prime}$ (RPB).

To experimentally determine the features of ion distributions in water, we set up surface sensitive X-ray diffraction experiments from a well behaved and controlled Langmuir monolayer at the air/water interface.  To extract the ion distributions, we employed the recently developed anomalous x-ray reflectivity technique for monolayers\cite{Vaknin2003}.  This basic type investigation has become feasible only with the advent of second generation synchrotron X-ray sources with novel insertion devices (i.e., undulator) and improved optics, which readily produce variable-energy X-ray beams with brilliancies capable of detecting a single atomic-layer even if not closely-packed. Another important advance in this regard is the development of liquid surface diffractometers first introduced by Als-Nielsen and Pershan\cite{Als-Nielsen1983}.

\begin{figure}[htl]
\includegraphics[width=2.4in]{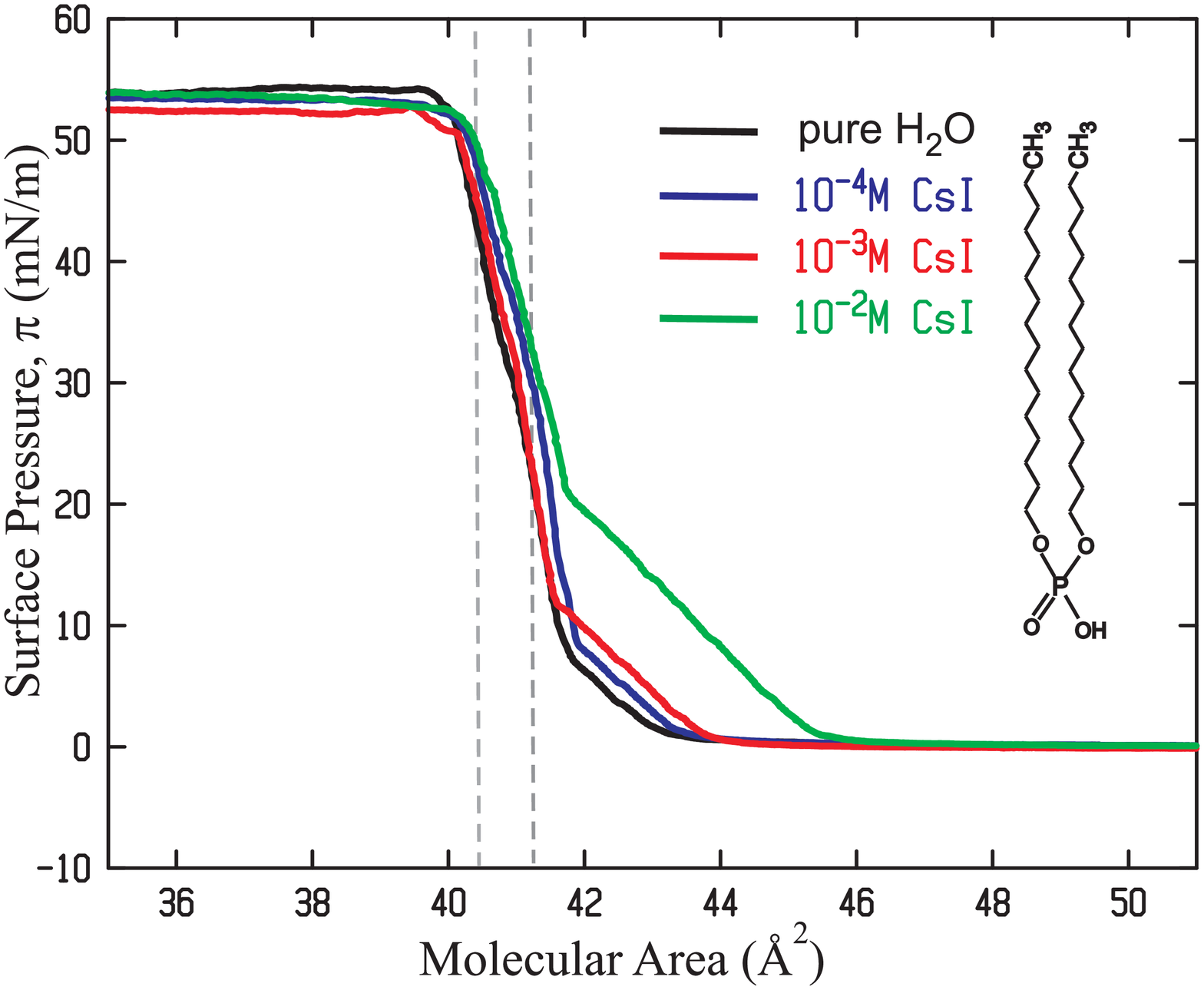}
\caption{\label{Iso1} (color online) Surface pressure versus molecular area for DHDP for different CsI concentrations as indicated. Reflectivity and GIXD were performed at constant surface-pressures 40mN/m and 30mN/m. The dashed lines indicate the region in isotherm reported in the present manuscript.}
\end{figure}

To manipulate ion bulk-concentrations, we used CsI (99.999\%, Sigma Corp. Cat\# 203033) solutions in ultra-pure water (experimental details handling monolayers for X-ray experiments are described in \cite{Vaknin2003b}), taking advantage of the $L_3$ resonance of Cs ions at 5.012 keV in anomalous reflectivity measurements.  To control surface charge density, monolayers of dihexadecyl-hydrogen-phosphate (DHDP, see Fig.\ \ref{Iso1}) (C$_{32}$H$_{67}$O$_{4}$P; MW = 546.86, Sigma, Corp. Cat\# D2631) were spread from 3:1 chloroform/methanol solutions at the air-water interface in a thermostated Langmuir trough\cite{Vaknin2003b}.  DHDP was chosen for this study, since it forms a simple in-plane structure at high enough surface pressures\cite{Gregory1997} and its hydrogen-phosphate head-group (R-PO$_4$H) has a pK$_a$ = 2.1, {\it presumably} guaranteeing almost complete dissociation [PO$_4^-$]/[R-PO$_4$H]
 $\approx$ 0.99999, with one electron-charge per molecule ($\sigma_0\approx$ 0.4 C/m$^2$).

X-ray reflectivity(XR) and grazing incident X-ray diffraction (GIXD) of monolayers at air/water interfaces were conducted on the Ames Laboratory Liquid Surface Diffractometer at the Advanced Photon Source (beam-line 6ID-B, described elsewhere\cite{Vaknin2001}) to determine the structure of the monolayer and ion distribution\cite{Vaknin2001,Als-Nielsen1989}.  Highly monochromatic X-ray beams (16.2 keV or 5.012 keV; $\lambda$ = 0.765334 and 2.47374 {\AA}, respectively) were selected by a downstream Si double-crystal monochromator, and deflected onto the liquid surface to a desired angle of incidence with respect to the liquid surface by a second monochromator (Ge(220)and Ge(111) crystals at 16.2 and 5.012 keV, respectively) located on the diffractometer. X-ray energy was calibrated with six different absorption edges to better than $\pm$3 eV and subsequent energy scans at fixed $Q_z$ in the course of the present study, accurately confirmed the $L_3$ energy of Cs$^+$ \cite{Bu-TBP}.  To extract the density profile across the interface from XR, a parameterized density profile $\rho(z) = \rho^{\prime}(z)+i\rho^{\prime\prime}(z)$ of the electron-density (ED) and the absorption-density (AD)(real and imaginary parts, respectively) is constructed by a sum of Error functions
\begin{equation}
\rho(z)=\frac{1}{2}\sum^{N}_{j=1}\mbox{Erf}\left(\frac{z-z_{j}} {\sqrt{2}\sigma_{j}}\right)(\rho_{j}-\rho_{j+1})+\frac{\rho_{N+1}}{2} \label{Erf}
\end{equation}
where $N$ is the number of interfaces, $\rho_{j} = \rho_{j}^\prime+i\rho{_j}^{\prime\prime}$; $\rho_{j}^\prime$ and $\rho{_j}^{\prime\prime}$ are the ED and AD of $j$th slab,  $z_{j}$ and $\sigma_{j}$ are the position and roughness of $j$th interface respectively, $\rho_{N+1}$ is the ED of the solution.  The AD profile is particularly important at the Cs resonance (5.012keV) as demonstrated below. The reflectivity is calculated using a recursive method\cite{Parratt1954} of the discretized density profile, Eq.\ (\ref{Erf}).\cite{Vaknin2003,Bu-TBP}.

Surface pressure versus molecular-area ($\pi-A$) isotherms of DHDP at various CsI salt concentrations ($n_b$) shown in Fig.\ \ref{Iso1}, were used to control surface-charge density, in particular, to identify conditions under which DHDP surface-density ($\sigma_0 = 1/A$) is independent of $n_b$.
For $\pi > 0$, the isotherm exhibits two distinct slopes, associated with crytalline tilted and non-tilted acyl-chains with respect to the surface normal as identified by GIXD and rod-scans\cite{Bu-TBP}. In the present study we focus on the non-tilted crystalline phase ($30 \lesssim\pi\lesssim 40$ \ mN/m), where the density variation at a fixed $\pi$ is less than 1.5\%.  This small variation in surface-density is corroborated by GIXD and rod-scan measurements of monolayers on CsI solutions in the 0.1-10$^{-5}$M range, which show two prominent inplane Bragg reflections consistent with the formation of 2D polycrystalline hexagonal symmetry\cite{Bu-TBP}.

\begin{figure}[htl]
\includegraphics[width=2.4 in]{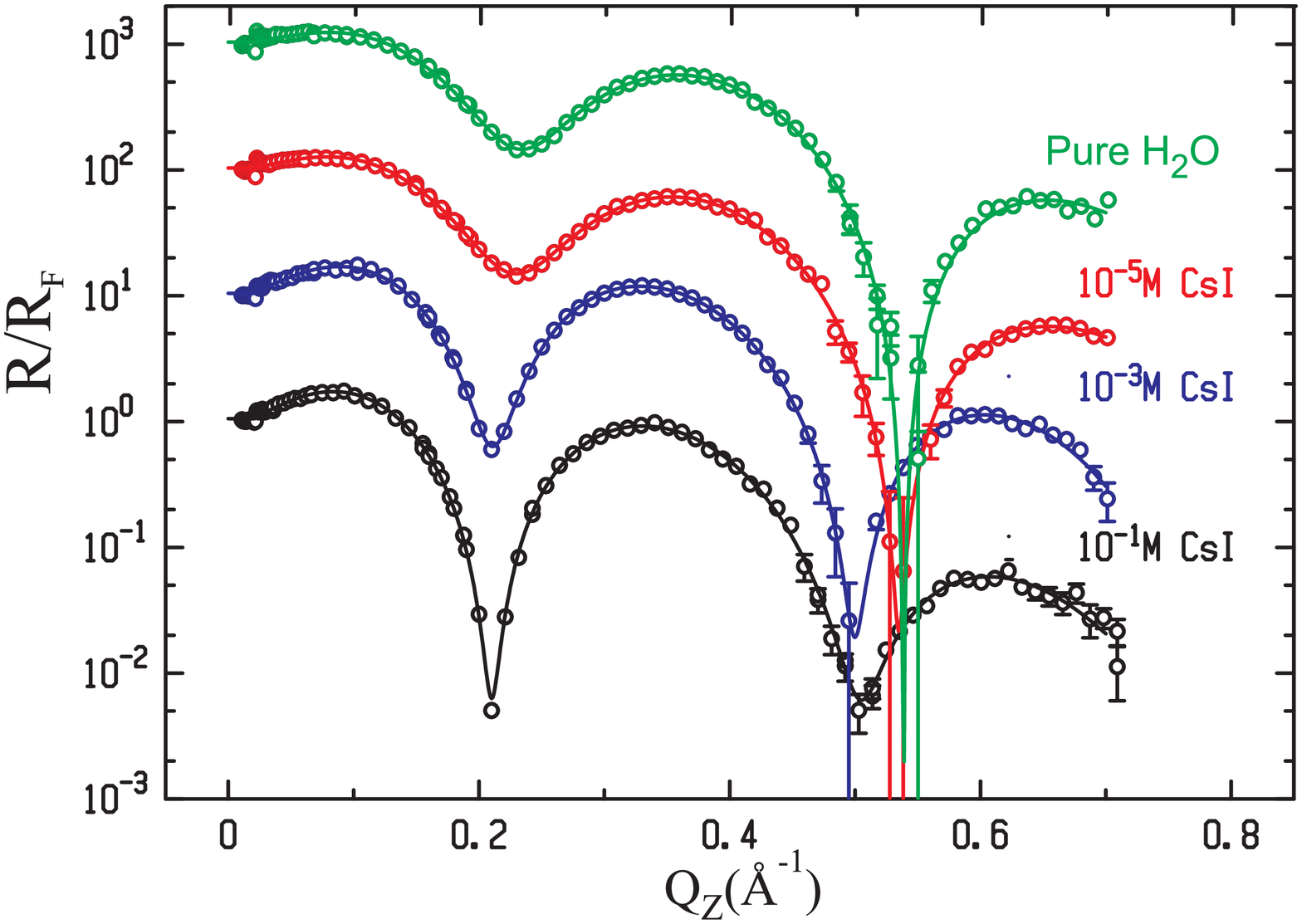}
\caption{\label{ref1} (color online) Measured normalizedXR curves and best fits (solid lines) for DHDP monolayers at various CsI concentrations at surface pressure 40mN/m (curves are shifted, by a decade each, for clarity).}
\end{figure}

Figure \ref{ref1} shows normalized reflectivity curves, $R/R_{F}$ (where $R_{F}$ is the calculated reflectivity of an ideally flat water interface), for DHDP ($\pi$ = 40mN/m) on pure H$_{2}$O, $10^{-5}$M, $10^{-3}$M, $10^{-1}$M, CsI measured at E=16.2keV.
All XR curves differ in the exact position and the sharpness of their minima, and the intensities of their maxima.  Similar reflectivity curves were obtained for $\pi = 30$ mN/m.  The solid lines are the best-fit calculated reflectivities based on refined density profiles that show differences mainly at and below the phosphate head-group region.  Since, the packing of DHDP is basically independent of salt concentration for $\pi$ = 40 mN/m, the reflectivity curves in Fig.\ \ref{ref1} qualitatively show a strong dependence of ion distribution at the interface on bulk ion concentration, in agreement with RPB discussed above.   To obtain counterion distributions, we combine the reflectivities at and off resonance (5.012 and 16.2 keV) into one data set and refine structural parameters using a space-filling model and applying volume constraints of different constituents\cite{Vaknin2003,Bu-TBP, Vaknin1991}.
\begin{figure}[htl]
\includegraphics[width=2.4 in]{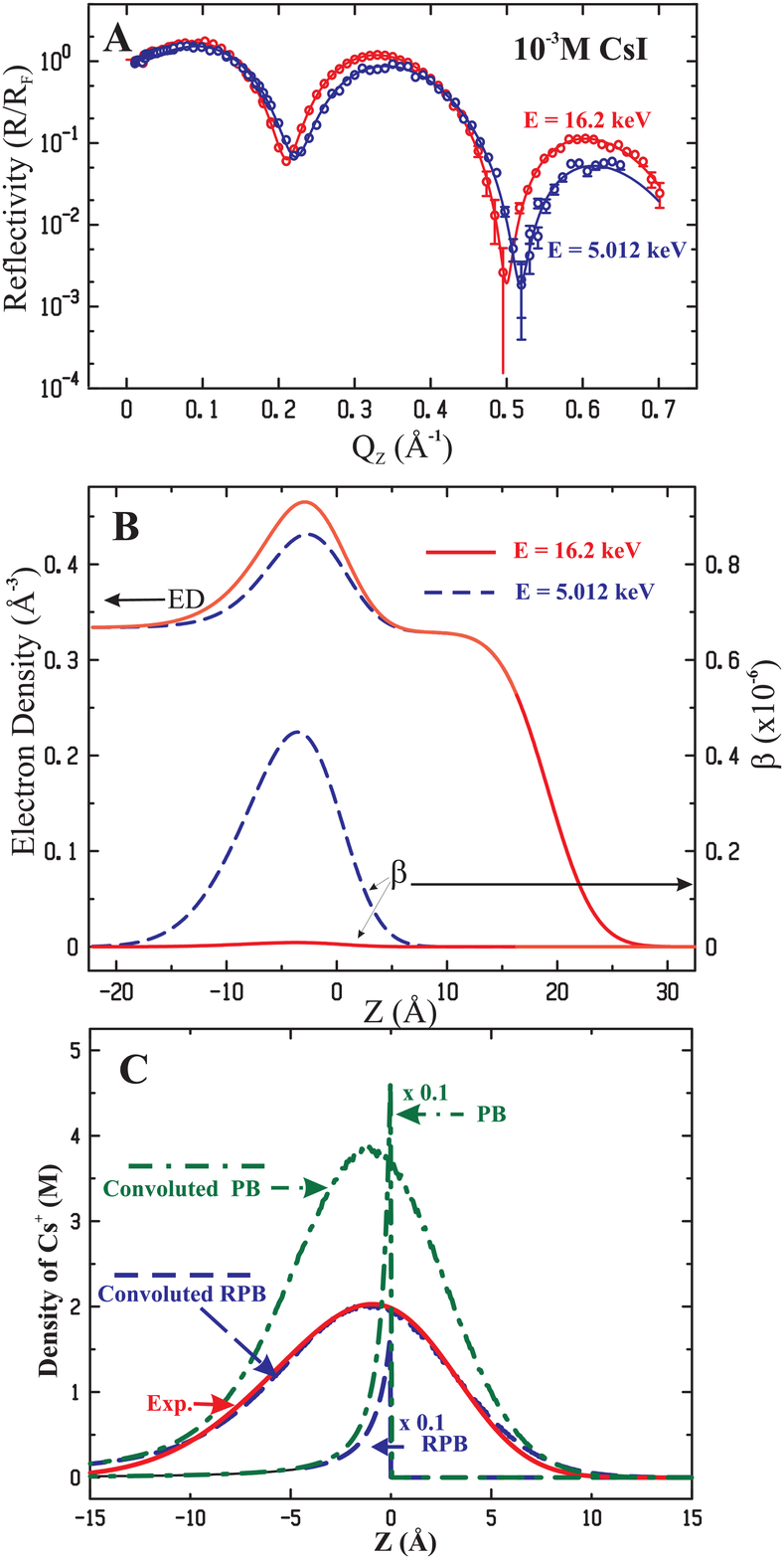}
\caption{\label{ref2} (color online) (A) Normalized X-ray reflectivities measured at 16.2 and 5.012 keV of DHDP monolayer spread on $10^{-3}$M CsI solution ($\pi$=40mN/m). Solid lines are calculated reflectivities using the ED and AD profiles shown in (B). The two data sets were combined and refined to a model with common structural adjustable parameters. (C) The solid line, obtained from the difference of the two ED's in (B), shows the experimental distribution of Cs$^+$, the dashed and dashed-dotted lines are calculated by RPB and PB respectively convoluted and non convoluted as indicated (non-convoluted calculations of PB and RPB are divided by 10).}
\end{figure}
Figure\ \ref{ref2}(A) shows reflectivities of DHDP spread on 10$^{-3}$M at $\pi$ = 40 mN/m at 16.2 and 5.012keV.  The solid lines are calculated from generalized density $\rho(z)$, obtained from parameter refinement of a single model-structure for the combined data sets, as shown in Fig.\ \ref{ref2}(B).  The AD curve for 5.012 keV up to a normalization factor is practically the profile of the counterions at the interface (there is a minute contribution to the AD from phosphorous in the head group region).  The difference between the ED's at and off resonance, normalized by $Z[1-f^{\prime}(E_{res})]$ where, $Z$ = 54 for Cs$^+$, gives the desired ionic distribution at the interface.  Figure\ \ref{ref2}(C) shows in a solid line the experimental Cs$^+$ distribution at the interface at 10$^{-3}$M.  Similar distributions, at other bulk CsI concentrations, are shown (solid lines) in Fig.\ \ref{pb2}(A).  Using the space-filling model to analyze the X-ray reflectivities off resonance,  and self consistently by integrating the distributions (obtained by the anomalous reflectivity) along the Z-axis, the number of counterions per DHDP was determined.
A compilation of the integrated number of ions at the interface is given in Fig.\ \ref{pb2}(B) (square symbols), along with the calculated RPB (solid line). The values from PB theory are also shown in Fig.\ \ref{pb2}(B)(triangles connected with a dashed line).
\begin{figure}
\includegraphics[width=2.4 in]{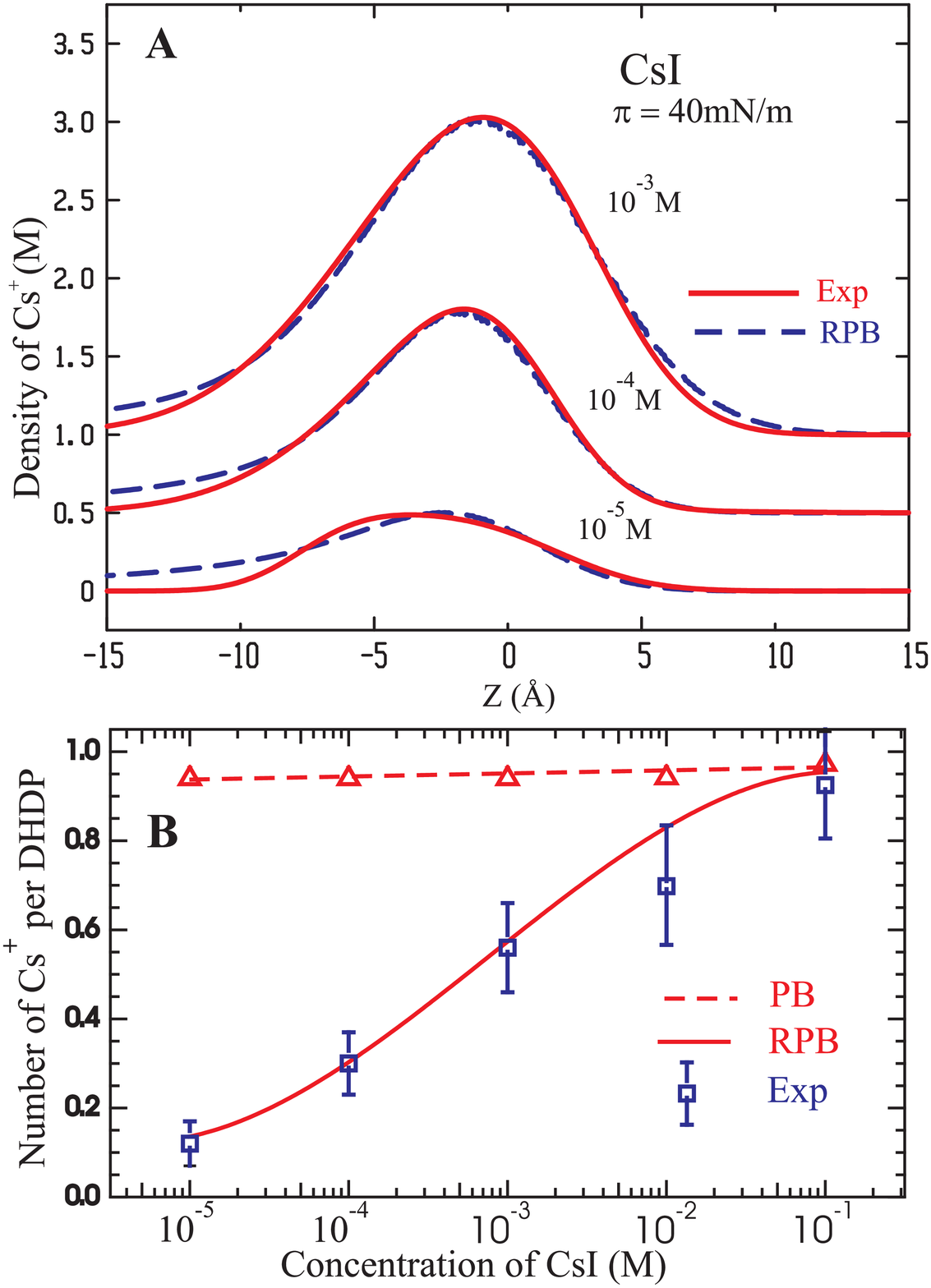}
\caption{\label{pb2} (color online) (A) Interfacial Cs$^+$ distributions (solid lines and shaded areas) determined from anomalous reflectivities as outlined in Fig.\ \ref{ref2} for various CsI bulk concentrations (shifted by 0.5M for clarity). Calculated distributions based on RPB are shown with dashed lines. (B) Number of Cs$^+$ ions per lipid ($\approx 41$\AA$^2$) (square symbols).  The dashed line is the PB integrated over the first 15\AA, and the solid line is obtained from RPB integrated over the same range.}
\end{figure}

To account for the fact that PB equation assumes point-like charges and an ideally flat interface, we propose to convolve the theoretical distribution $n^+(z)$ with a Gaussian function as follows,
\begin{equation}
n_r^+(z) = \frac{1}{\Gamma\sqrt{2\pi}}\int{n^+(z^{\prime})\text{e}^{-\frac{(z-z^{\prime})^2}{2\Gamma^2}}\text{dz}^{\prime}} \label{conv}
\end{equation}
The convolution function is justified by the following assumptions: 1) interface fluctuations are Gaussian and 2) the wavelength of the capillary waves at the interface are larger than molecular size. The width of the Gaussian $\Gamma$ is determined by surface roughness, which is independently determined from the reflectivity, and therefore the convolution does not involve any new parameters. The calculated ionic distribution using RPB for $n_b$ = 10$^{-3}$M and its convolution are shown as dashed lines in Fig.\ \ref{ref2}(C) ($\Gamma = 3.9 $ \AA \ is determined from our reflectivity experiments, and using pH-pKa = 2.5, within the range of uncertainty of the measured pH of our pure water), superimposed on our experimental data (solid line) with no adjustable parameters. Figure\ \ref{pb2}(A) shows counterion distributions for three different $n_b$'s (solid lines), with corresponding calculations of RPB and convoluted as in Eq.\ (\ref{conv}) (dashed lines). The agreement between RPB theory and experiment with a simple smearing of the distribution, Eq.\ (\ref{conv}) is remarkably good, except for slight deviations at distances $Z \lesssim -10 $ \AA \ away from the interface.\cite{Bu-TBP}.

In this study, we have shown remarkable agreement of the RPB theory with experiment.  We obtained both the distribution and the integrated number of monovalent ions per charge at the interface over five orders of magnitude in ion-bulk concentrations. 

Our experimental results for the ion distribution are entirely consistent with water being described with a continuum of bulk dielectric constant.  Corrections due to finite size ionic radius, charge modulations, short-range interactions, image charges or water restructuring were not necessary for describing the experimental data, implying that such effects, if relevant, would change the distribution at distances shorter than $\sim$ 3 {\AA}\cite{comment1}.
The fact that the pK$_a$ of the amphiphiles in our experiment is one of the lowest available also shows the dramatic effects that the renormalization Eq.\ (\ref{sigma}) (RPB) has at high surface charges. The understanding we gained with the monovalent ions is of critical importance to our ongoing investigations of charged interfaces at multivalent-ion solutions.

We thank D. S. Robinson and D. Wermeille for technical support at the 6-ID beamline, to M. Gordon for helpful discussions, and the Referee for guiding comments. The MUCAT sector at the APS is supported by the U.S. DOE Basic Energy Sciences, Office of Science, through Ames Laboratory under contract no. W-7405-Eng-82. The work at Ames Laboratory is supported by the U.S. DOE, Basic Energy Sciences, Office of Science, under contract no. W-31-109-Eng-38, the work of AT is partially supported by NSF-DMR-0426597.


\begin{references}
\bibitem{comment1}As pointed out to us by the Referee, the addition of any function  $F(z)$ to the calculated RPB distribution with the following properties: a) $\int_0^\Gamma F(z)dz \sim 0$ and b) $max(|F(z)|) \lessapprox max(ED_{Cs}(z))$, where $ED_{Cs}(z)$ is the experimentally determined distribution,  leaves the convolution invariant.
\end{references}
\end{document}